\documentclass{emulateapj}
\usepackage{apjfonts}

\def\chandra{{\it Chandra\/}}

\def\spitzer{{\it Spitzer\/}}

\def\xray{\hbox{X-ray}}
\def\ecdfs{\hbox{E-CDF-S}}
\def\cdfs{\hbox{CDF-S}}
\def\cdfn{\hbox{CDF-N}}
\def\etal{{et\,al.}}

\def\ltsima{$\; \buildrel < \over \sim \;$}
\def\simlt{\lower.5ex\hbox{\ltsima}}
\def\gtsima{$\; \buildrel > \over \sim \;$}
\def\simgt{\lower.5ex\hbox{\gtsima}}
\def\kms{\ifmmode{~{\rm km~s^{-1}}}\else{~km s$^{-1}$}\fi}
\def\lsim{\lower0.3em\hbox{$\,\buildrel <\over\sim\,$}}
\def\gsim{\lower0.3em\hbox{$\,\buildrel >\over\sim\,$}}

\def\msol{$M_\odot$}
\def\h2{H$_2$}
\def\flux{ergs~cm$^{-2}$~s$^{-1}$}
\def\xlum{ergs~s$^{-1}$}

\def\arcmin{\mbox{$^\prime$}}
\def\sfr{$M_{\odot}$ yr$^{-1}$}

\def\aap{A\&A}
\def\apj{ApJ}
\def\apjl{ApJL}
\def\apjs{ApJS}
\def\aj{AJ}
\def\mnras{MNRAS}
\def\araa{ARA\&A}

\def\nat{Nature}

\begin{document}

\shortauthors{\large LEHMER ET AL.}
\shorttitle{\large AGN ACTIVITY IN THE SSA22 PROTOCLUSTER}

%
\title{The {\itshape Chandra} Deep Protocluster Survey: Evidence for an Enhancement of AGN Activity in the SSA22 Protocluster at \lowercase{$z$} = 3.09}
%

\author{
B.~D.~Lehmer,$^1$
D.~M.~Alexander,$^1$
J.~E.~Geach,$^1$
Ian~Smail,$^{1,2}$
A.~Basu-Zych,$^3$
F.~E.~Bauer,$^3$
S.~C.~Chapman,$^4$
Y.~Matsuda,$^5$
C.~A.~Scharf,$^3$
M.~Volonteri,$^6$
\& T.~Yamada$^5$\\
}

\altaffiltext{1}{Department of Physics, Durham University, South Road, Durham, DH1 3LE, UK}
\altaffiltext{2}{Institute for Computational Cosmology, Durham University, South Road, Durham, DH1 3LE, UK}
\altaffiltext{3}{Columbia Astrophysics Laboratory, Columbia University, Pupin Labortories, 550 W. 120th St., Rm 1418, New York, NY 10027, USA}
\altaffiltext{4}{Institute of Astronomy, Madingley Road, Cambridge CB3 0HA, UK}
\altaffiltext{5}{National Astronomical Observatory of Japan, Tokyo 181-8588, Japan}
\altaffiltext{6}{Department of Astronomy, University of Michigan, Ann Arbor, MI, USA}

%
\begin{abstract}
%

We present results from a new ultra-deep $\approx$400~ks \chandra\
observation of the SSA22 protocluster at $z=3.09$.  We have studied the \xray\
properties of 234 $z \sim 3$ Lyman break galaxies (LBGs; protocluster and
field) and 158 $z = 3.09$ Ly$\alpha$ emitters (LAEs) in SSA22 to measure the
influence of the high-density protocluster environment on the accretion
activity of supermassive black holes (SMBHs) in these UV-selected star forming
populations.  We detect individually \xray--emission from active galactic
nuclei (AGNs) in six LBGs and five LAEs; due to small overlap between the LBG
and LAE source population, ten of these sources are unique.  At least six and
potentially eight of these sources are members of the protocluster.  These
sources have rest-frame \hbox{8--32~keV} luminosities in the range of
\hbox{$L_{\rm 8-32~keV} =$~(3--50)~$\times 10^{43}$~\xlum} and an average
observed-frame \hbox{2--8~keV} to \hbox{0.5--2~keV} band-ratio of $\approx$0.8
(mean effective photon index of $\Gamma_{\rm eff} \approx$~1.1), suggesting
significant absorption columns of \hbox{$N_{\rm H} \simgt
10^{22}$--$10^{24}$~cm$^{-2}$}.  We find that the fraction of LBGs and LAEs in
the $z = 3.09$ protocluster harboring an AGN with \hbox{$L_{\rm 8-32~keV}
\simgt 3 \times 10^{43}$~\xlum} is $9.5^{+12.7}_{-6.1}$\% and
$5.1^{+6.8}_{-3.3}$\%, respectively.  These AGN fractions are somewhat larger
(by a mean factor of $6.1^{+10.3}_{-3.6}$; significant at the $\approx$95\%
confidence level) than $z \sim 3$ sources found in lower-density ``field''
environments.  Theoretical models imply that these results may be due to the
presence of more actively growing and/or massive SMBHs in LBGs and LAEs within
the protocluster compared to the field.  Such a result is expected in a
scenario where enhanced merger activity in the protocluster drives accelerated
galaxy and SMBH growth at \hbox{$z \simgt$~2--3}.  Using \spitzer\ IRAC imaging
we found that the fraction of IRAC detected LBGs is significantly larger in the
protocluster than in the field (by a factor of 3.0$^{+2.0}_{-1.3}$).  From
these data, we constrained the median rest-frame $H$-band luminosity in the
protocluster to be \hbox{$\simgt$1.2--1.8} times larger than that for the field.  When
combined with our \xray\ data, this suggests that both galaxies and SMBHs grew
more rapidly in protocluster environments.

%
\end{abstract}
%


%
\section{Introduction}
%

Over the last decade, investigations have revealed that galaxies with spheroid
components (i.e., elliptical galaxies, lenticulars, and spiral galaxy bulges)
ubiquitously contain supermassive black holes (SMBHs) in their cores (e.g.,
Kormendy \& Richstone 1995; Magorrian \etal\ 1998).  These studies have also
confirmed the existence of a tight relationship between the mass of the SMBHs
and the stellar mass of the spheroid, suggesting a causal connection between
the growth of these two galactic components (e.g., Gebhardt \etal\ 2000).

In addition, theories of large-scale structure formation in a CDM Universe predict
that galaxy formation is accelerated in high-density regions (Kauffmann~1996;
de~Lucia \etal\ 2006).  Observational studies have provided convincing support
for this hypothesis, showing that there is a strong relationship between galaxy
stellar age and local environment in the nearby universe (e.g., Smith \etal\
2008 and references therein); the most evolved and massive galaxies reside in
the highest density regions of local clusters, while more typical galaxies that
are undergoing significant star formation are generally found in lower density
environments (e.g., Lewis \etal\ 2002).  Studies of distant galaxy populations
at $z \approx$~1 are finding that the star-formation activity occurs in higher
density environments than seen locally (e.g., Geach \etal\ 2006; Elbaz \etal\
2007; Heinis \etal\ 2007; Cooper \etal\ 2008; Poggianti \etal\ 2008).  These
studies indicate that a reversal in the star-formation rate (SFR)/galaxy density
relation occurs at $z \simgt 1$, where the most intense galaxy growth is
expected to occur in the highest density clusters.

The progenitors of the highest density clusters in the local universe are also
expected to be the highest density structures at $z \simgt$~2--3 and should be
undergoing vigorous star formation during their assemblage (e.g., Governato
\etal\ 1998).  These protoclusters are identified through overdense redshift
``spikes'' in high-redshift galaxy surveys of blank fields (e.g., Adelberger
\etal\ 1998; Steidel \etal\ 2003) and in the vicinity of certain powerful radio
galaxies (e.g., Venemans \etal\ 2007).  It is therefore plausible to expect that
if the growth of galaxies and their central SMBHs are causally linked, then the
highest density structures will also be the sites of significant SMBH
accretion, identifiable as active galactic nuclei (AGNs).

To detect and study typical AGN ($L_{\rm X} \simgt 10^{43-44}$~\xlum) in $z
\simgt$~2--3 protocluster galaxies requires significant optical spectroscopic
and \xray\ observational investments.  Therefore, few programs have yet
explored how AGN activity varies as a function of environment at these
redshifts.  Nonetheless, initial studies of high-density regions and clusters
in the \hbox{$z \simgt$~0.5--3} universe have provided suggestive evidence for
an elevation in the AGN activity in such high-density environments compared to
the field (see, e.g., Pentericci \etal\ 2002; Gilli \etal\ 2003; Johnson \etal\
2003; Smail \etal\ 2003; Croft \etal\ 2005; Eastman \etal\ 2007; Silverman
\etal\ 2008).  However, a rigorous quantification of such an enhancement in AGN
activity has yet to be obtained for actively forming protoclusters that are
precursors to rich clusters at $z = 0$.

The $z = 3.09$ SSA22 protocluster was originally identified by Steidel \etal\
(1998) as a significantly overdense region (by a factor of $\approx$4--6)
through spectroscopic follow-up observations of $z \sim 3$ candidate Lyman
break galaxies (LBGs).  Theoretical modelling indicates that the protocluster
should collapse into a $z = 0$ structure resembling a rich local cluster with a
total mass $\simgt$$10^{15}$~\msol\ (e.g., Coma; see Steidel \etal\ 1998 for
details).  Since its discovery, the protocluster has been found to contain a
factor of $\approx$6 overdensity of Ly$\alpha$ emitters (LAEs; Steidel \etal\
2000; Hayashino \etal\ 2004; Matsuda \etal\ 2005) and several remarkable bright
extended Ly$\alpha$-emitting blobs (LABs; Steidel \etal\ 2000; Matsuda \etal\
2004), which are hypothesized to be sites of either cooling flows or
starburst/AGN outflows (e.g., Bower \etal\ 2004; Geach \etal\ 2005, Wilman
\etal\ 2005; Geach \etal\ in preparation).  Therefore, SSA22 is an ideal
field for studying how SMBH growth depends on environment in the $z
\simgt$~2--3 universe.

%
%
\begin{figure}
\centerline{
\includegraphics[width=9cm]{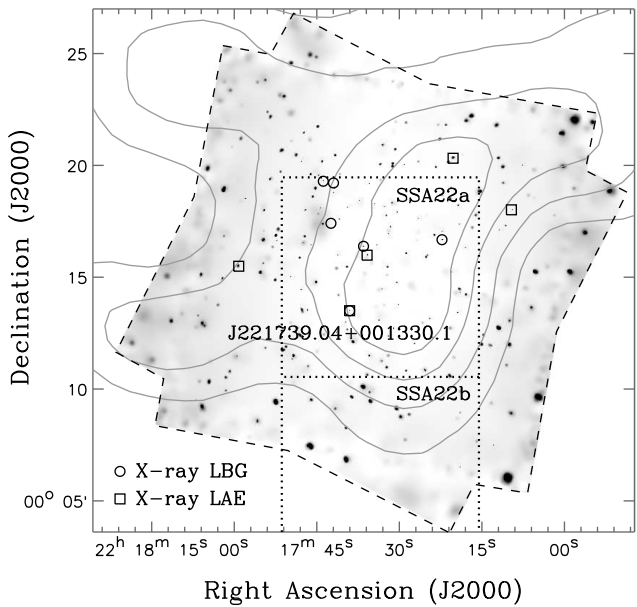}
}
\caption{
Adaptively-smoothed \hbox{0.5--8~keV} \chandra\ image (with boundaries shown as
a {\it dashed polygon\/}) of the SSA22 field.  X-ray detected LBGs and LAEs
have been highlighted with open circles and squares, respectively.  The dotted
regions show the extent of the Steidel \etal\ (2003) ``SSA22a'' and ``SSA22b''
LBG surveys.  We note that the Hayashino \etal\ (2004) sample of LAEs covers a
region larger than that shown here.  LAE source-density contours are shown in
gray (computed using our LAE catalog and a spatially-varying density extraction
circle with radius of 3\farcm0) and have levels of 1.0, 1.3, 1.7, and 2.0 LAEs
arcmin$^{-2}$.  For comparison the LAE source-density in the \ecdfs\ field
comparison sample is $\approx$0.3 LAEs arcmin$^{-2}$.  We note that most of the
\xray\ sources lie in regions of high LAE density.  For reference, we have
highlighted the location of \hbox{J221739.04+001330.1}, which is both an LBG
and LAE.  
}
\end{figure}

In this paper, we utilize new ultra-deep $\approx$400~ks \chandra\ observations
of the $z = 3.09$ SSA22 protocluster region to identify luminous AGNs
indicative of accreting SMBHs.  We use this data to place constraints on how
the AGN properties (e.g., luminosity and \xray\ spectra) and frequency in the
protocluster compares with AGNs identified in lower density field regions of
the \chandra\ Deep Field-North (\cdfn) and Extended \chandra\ Deep Field-South
(\ecdfs).  The Galactic column density for SSA22 is \hbox{$4.6 \times
10^{20}$~cm$^{-2}$} (Stark \etal\ 1992).  All \xray\ fluxes and luminosities
quoted in this paper have been corrected for Galactic absorption.  The
coordinates throughout this paper are J2000.0.  \hbox{$H_0$ = 70~km s$^{-1}$
Mpc$^{-1}$}, $\Omega_{\rm M}$ = 0.3, and $\Omega_{\Lambda}$ = 0.7 are adopted
(e.g., Spergel \etal\ 2003), which give the age of the Universe as 13.5~Gyr and
imply a $z=3.09$ look-back time and spatial scale of 11.4~Gyr and
7.6~kpc~arcsec$^{-1}$, respectively.

\section{Chandra Observations}

We obtained a $\approx$400~ks \chandra\ exposure consisting of four
\hbox{$16.9\arcmin \times 16.9\arcmin$} ACIS-I pointings (\chandra\ Obs-IDs
8034, 8035, 8036, and 9717 [taken between 01~Oct~2007 and 30~Dec~2007]; PI:
D.~M.~Alexander) centered on the SSA22a LBG region surveyed by Steidel \etal\
(2003; see Fig.~1).\footnote{We note that the $\approx$79~ks \hbox{ACIS-S}
exposure (Obs-ID 1694 [taken on 10~Jul~2001]; PI: G.~P.~Garmire) was not merged
with our \hbox{ACIS-I} observations since, due to non negligible differences in
aim points and backgrounds, it did not improve data quality or the number of
\chandra\ source detections.}  We note that the SSA22a region represents a
high-density pocket of the whole SSA22 protocluster at \hbox{$z =$~3.06--3.12},
which is known to extend to scales much larger ($\simgt$1~deg$^2$) than the
SSA22a LBG survey region as well as the region covered by our \chandra\
observations (Yamada \etal\ in preparation).  Due to small variations in the
aim points and roll angles of the observations, the total exposure covers a
solid angle of $\approx$330~arcmin$^2$ with more than 60\% of the field
reaching effective vignetting-corrected exposures $\simgt$300~ks.  

\chandra\ \hbox{X-ray} Center (hereafter CXC) pipeline software was used for
basic data processing, and the pipeline version 7.6.11 was used in all
observations.  The reduction and analysis of the data used \chandra\
Interactive Analysis of Observations ({\ttfamily CIAO}) Version~3.4 tools
whenever possible;\footnote{See http://cxc.harvard.edu/ciao/ for details on
{\ttfamily CIAO}.} however, custom software, including the Tools for ACIS
Real-time Analysis ({\ttfamily TARA}; Broos \etal\ 2000) software
package,\footnote{{\ttfamily TARA} is available at
http://www.astro.psu.edu/xray/docs.} were also used extensively.  The details
of our data processing procedure have been outlined in $\S$~2.2 of Luo \etal\
(2008).

We have compiled \chandra\ point-source catalogs, which were generated by
running {\ttfamily wavdetect} at a false-positive probability threshold of
10$^{-5}$ over three standard bandpasses: \hbox{0.5--8~keV} (full band; FB),
\hbox{0.5--2~keV} (soft band; SB), and \hbox{2--8~keV} (hard band; HB).  The
significance of each source was then individually analyzed using the {\ttfamily
ACIS EXTRACT} software package (Broos \etal\ 2002) and our source list was then
filtered to include only sources that had high statistical probabilities
($\simgt$99\% confidence) of being true sources considering their local
backgrounds.  

In total, our final point-source catalog contained 297
\hbox{X-ray} point sources over the entire field, and these sources were used
in our analysis below.  The survey reaches ultimate sensitivity limits of
$\approx$4.8~$\times 10^{-17}$~\flux\ and $\approx$2.7~$\times 10^{-16}$~\flux\
for the \hbox{0.5--2~keV} and \hbox{2--8~keV} bands, respectively; at $z = 3$,
these limits correspond to rest-frame \hbox{2--8~keV} and \hbox{8--32~keV}
luminosities of $\approx$3.7~$\times 10^{42}$~\xlum\ and $\approx$2.1~$\times
10^{43}$~\xlum, respectively.  Number count analyses show that the \chandra\
source density for the SSA22 field is consistent with the \cdfn\ and \cdfs\
surveys (see, e.g., Alexander \etal\ 2003; Bauer \etal\ 2004; Luo \etal\ 2008).

In a subsequent paper (Lehmer \etal\ in preparation) we will provide
full details regarding our \chandra\ data analyses and point-source catalog
production; data products and \chandra\ point-source catalogs will be made
publicly available.  

%
%
\begin{table*}
\begin{minipage}{180mm}
\begin{center}
\caption{Properties of $z\sim3$ X-ray Detected Sources in the SSA22 and CDF samples}
\begin{tabular}{lcccccccc}
\hline\hline
 & Source Name & $\theta$ & & Band Ratio & & $L_{\rm 2-8~keV}$ & $L_{\rm 8-32~keV}$ & Optical \\ 
\multicolumn{1}{c}{Survey Field} & (J2000.0) & (arcsec) & $z$ & (2--8~keV)/(0.5--2~keV) & $\Gamma_{\rm eff}$ & ($\log$ ergs s$^{-1}$) & ($\log$ ergs s$^{-1}$) & Classification \\
 \multicolumn{1}{c}{(1)} & (2) & (3) & (4) & (5) & (6) & (7) & (8) & (9) \\
\hline
\multicolumn{9}{c}{\xray--Detected Lyman Break Galaxies ($z \approx$~2--3.4)} \\
\hline
    SSA22 \ldots\ldots\ldots\ldots\ldots\ldots\ldots &      J221722.25+001640.6 &                      0.67 &                     3.353 &    0.78$^{+0.17}_{-0.15}$ &    1.13$^{+0.18}_{-0.17}$ &   44.07$^{+0.06}_{-0.06}$ &   44.58$^{+0.06}_{-0.07}$  &             QSO \\
\vspace{0.01in}
                                                     &      J221736.51+001622.9 &                      0.52 &                     3.084 &    0.80$^{+0.30}_{-0.24}$ &    1.11$^{+0.30}_{-0.29}$ &   43.51$^{+0.10}_{-0.10}$ &   44.03$^{+0.11}_{-0.11}$  &             QSO \\
\vspace{0.01in}
                                                     &      J221739.04+001330.1 &                      0.91 &                     3.091 &                   $>$1.83 &                   $<$0.42 &                  $<$43.08 &   44.01$^{+0.13}_{-0.14}$  &   Gal, LAB2$^a$ \\
\vspace{0.01in}
                                                     &      J221741.97+001912.8 &                      0.58 &          $\approx$3$^{b}$ &                $\approx$1 &                   1.4$^c$ &                  $<$43.12 &                  $<$43.73  &                 \\
\vspace{0.01in}
                                                     &      J221742.43+001724.6 &                      0.87 &          $\approx$3$^{b}$ &                   $<$1.48 &                   $>$0.60 &   42.82$^{+0.23}_{-0.26}$ &                  $<$43.59  &                 \\
\vspace{0.01in}
                                                     &      J221743.82+001917.4 &                      0.32 &                     2.857 &    1.33$^{+0.57}_{-0.44}$ &    0.68$^{+0.31}_{-0.30}$ &   43.31$^{+0.12}_{-0.12}$ &   44.10$^{+0.10}_{-0.11}$  &             Gal \\
\vspace{0.01in}
\\
    CDF-N \ldots\ldots\ldots\ldots\ldots\ldots\ldots &      J123622.63+621306.4 &                      0.47 &                     2.981 &    0.76$^{+0.42}_{-0.32}$ &    1.17$^{+0.50}_{-0.40}$ &   42.66$^{+0.12}_{-0.13}$ &   43.25$^{+0.15}_{-0.19}$  &             Gal \\
                                                     &      J123633.53+621418.3 &                      0.35 &                     3.413 &    0.41$^{+0.08}_{-0.07}$ &    1.74$^{+0.16}_{-0.16}$ &   43.61$^{+0.04}_{-0.04}$ &   43.86$^{+0.06}_{-0.07}$  &             QSO \\
                                                     &      J123644.11+621311.2 &                      0.87 &                     2.929 &                $\approx$1 &                   1.4$^c$ &                  $<$42.29 &                  $<$43.01  &             Gal \\
                                                     &      J123651.56+621042.2 &                      0.91 &                     2.975 &                   $<$1.31 &                   $>$0.68 &   42.23$^{+0.14}_{-0.21}$ &                  $<$43.01  &             Gal \\
                                                     &      J123655.77+621201.1 &                      0.29 &          $\approx$3$^{b}$ &    0.48$^{+0.17}_{-0.14}$ &    1.60$^{+0.32}_{-0.28}$ &   42.99$^{+0.07}_{-0.07}$ &   43.32$^{+0.12}_{-0.13}$  &                 \\
                                                     &      J123702.58+621244.3 &                      0.54 &          $\approx$3$^{b}$ &    0.39$^{+0.11}_{-0.09}$ &    1.78$^{+0.24}_{-0.22}$ &   43.25$^{+0.05}_{-0.05}$ &   43.47$^{+0.09}_{-0.10}$  &                 \\
                                                     &      J123704.31+621446.5 &                      0.13 &                     2.211 &                   $<$1.93 &                   $>$0.33 &   42.03$^{+0.15}_{-0.24}$ &                  $<$42.98  &             AGN \\
                                                     &      J123719.88+620955.2 &                      0.16 &                     2.647 &    0.71$^{+0.08}_{-0.08}$ &    1.24$^{+0.10}_{-0.10}$ &   43.64$^{+0.03}_{-0.03}$ &   44.19$^{+0.04}_{-0.04}$  &             AGN \\
\hline
\multicolumn{9}{c}{\xray--Detected Ly$\alpha$ Emitters ($z = 3.1$)} \\
\hline
    SSA22 \ldots\ldots\ldots\ldots\ldots\ldots\ldots &      J221709.64+001800.7 &                      0.91 &                     3.106 &                   $<$0.86 &                   $>$1.05 &   43.52$^{+0.12}_{-0.13}$ &                  $<$44.06  &             Gal \\
\vspace{0.01in}
                                                     &      J221720.24+002019.1 &                      0.13 &                     3.105 &    0.42$^{+0.10}_{-0.09}$ &    1.68$^{+0.21}_{-0.20}$ &   44.20$^{+0.05}_{-0.05}$ &   44.35$^{+0.08}_{-0.09}$  &             Gal \\
\vspace{0.01in}
                                                     &      J221735.86+001559.1 &                      0.34 &                     3.094 &    0.78$^{+0.26}_{-0.21}$ &    1.13$^{+0.26}_{-0.25}$ &   43.62$^{+0.08}_{-0.08}$ &   44.13$^{+0.10}_{-0.10}$  &      Gal, LAB14 \\
\vspace{0.01in}
                                                     &      J221739.04+001330.1 &                      0.91 &                     3.091 &                   $>$1.83 &                   $<$0.42 &                  $<$43.08 &   44.01$^{+0.13}_{-0.14}$  &   Gal, LAB2$^a$ \\
\vspace{0.01in}
                                                     &      J221759.19+001529.4 &                      0.69 &                     3.096 &    0.66$^{+0.23}_{-0.19}$ &    1.28$^{+0.30}_{-0.28}$ &   43.75$^{+0.08}_{-0.08}$ &   44.17$^{+0.11}_{-0.12}$  &       Gal, LAB3 \\
\vspace{0.01in}
\\
        E-CDF-S \ldots\ldots\ldots\ldots\ldots\ldots &    J033307.61$-$275127.0 &                      0.00 &                       1.6 &    0.42$^{+0.03}_{-0.03}$ &    1.67$^{+0.07}_{-0.07}$ &   44.58$^{+0.02}_{-0.02}$ &   44.78$^{+0.03}_{-0.03}$  &             AGN \\
                                                     &    J033316.86$-$280105.2 &                      0.40 &                       3.1 &                   $>$2.48 &                   $<$0.04 &                  $<$43.25 &   44.43$^{+0.11}_{-0.12}$  &             AGN \\
\hline
\end{tabular}
\end{center}
\indent NOTES.--- Col.(1): \chandra\ survey field where each source was detected. Col.(2): \chandra\ source name.  Col.(3): Matching offset between optical and \chandra\ source positions in arcseconds.  Col.(4): Best available redshift for each source.  Col.(5): Observed-frame \hbox{2--8~keV} to \hbox{0.5--2~keV} band ratio.  Col.(6): Inferred effective photon index ($\Gamma_{\rm eff}$). Col.(7)--(8): Logarithm of the rest-frame \hbox{2--8~keV} and \hbox{8--32~keV} luminosity in units of ergs s$^{-1}$. Col.(9): Notes on optical spectroscopic source types from Steidel \etal\ (2003) for the SSA22 and \cdfn\ LBGs, Matsuda \etal\ (2005) for the SSA22 LAEs and Gronwall \etal\ (2007) for the \ecdfs\ LAEs.  We have also noted known LAEs that host Ly$\alpha$--emitting blobs (LABs; see Geach \etal\ in preparation).
\vspace{0.1in} \\
$^a$ Denotes duplicate LBG and LAE. \\
$^b$ Redshift of $z = 3$ was assumed for all LBG candidates that did not have a spectroscopic counterpart. \\
$^c$ Sources that were detected in only the \hbox{0.5--8~keV} bandpass were assumed to have an effective photon-index of $\Gamma_{\rm eff} = 1.4$.\\
\end{minipage}
\end{table*}

\section{Sample Generation and X-ray Matching}

\subsection{Selection of SSA22 and Field Comparison Samples}

We began by assembling samples of $z\sim3$ LBGs and LAEs that lie within the
\chandra-observed region of the SSA22 protocluster (hereafter, SSA22 samples).
For the LBG sample, we utilized the ``SSA22a'' and ``SSA22b'' $U_n$-dropout
source list provided by Steidel \etal\ (2003).  In total, 234 LBGs lie within
the $\approx$400~ks \chandra\ SSA22 observations in an $\approx$122~arcmin$^2$
region (see Fig.~1 and Lehmer \etal\ in preparation).  These LBGs have $\cal
R$-band magnitudes ranging from 20.8 to 25.6 (median \hbox{${\cal R} = 24.7$}).
Of the 234 SSA22 LBGs, 107 ($\approx$46\%) have spectroscopic redshifts from
Steidel \etal\ (2003), 27 ($\approx$25\%) of which are within the redshift
bounds \hbox{$z =$~3.06--3.12}, which we consider to be members of the
protocluster (see Matsuda \etal\ 2005 for justification).  We note that the
redshift range of the LBG sample in general is \hbox{$z \approx$~2.0--3.4} and
therefore we cannot conclude whether sources without spectroscopic redshifts
are inside or out of the protocluster redshift spike.  For the LAE sample, we
utilized 158 $z \approx 3.09$ LAEs from Hayashino \etal\ (2004) that were
within the extent of the \chandra\ observations ($\approx$292~arcmin$^2$
overlap) and had observed-frame Ly$\alpha$ equivalent widths of EW$_{\rm
obs}$~$>$~80~\AA\ and narrow-band 5000~\AA\ magnitudes brighter than NB~$<
25.4$ (AB).  While it is possible that some of the LAEs in our sample may be
low-redshift interlopers, spectroscopic follow-up of a larger and less
conservative sample of 118 out of 271 LAEs with NB~$< 25.4$ and EW$_{\rm obs}
\simgt$~69~\AA\ find only two contaminating sources at $z = 0.332$ and 0.329,
which were found to be [O~{\small II}] $\lambda$3737 doublet emitters
(Y.~Matsuda \& T.~Yamada private communication; see also, Matsuda \etal\ 2005
for further detail).  Therefore, we expect that $\simgt$97\% of the LAEs in our
sample are indeed located in the protocluster.  We further note that seven
($\approx$26\%) of our LBGs with spectroscopic redshifts in the range of
\hbox{$z =$~3.06--3.12} are also LAEs.  In total, our SSA22 samples contain 384
unique sources at $z \sim 3$.

For the purpose of comparing the accretion properties of our SSA22 sample with
LBGs and LAEs found in more typical low-density regions of the $z\sim3$
universe, we created \chandra\ Deep Field (CDF) samples of LBGs in the \cdfn\
and LAEs in the \ecdfs\ (hereafter, field comparison samples).  For our field
LBG sample, we used the 146 ``HDF-N'' $U_n$-dropouts from Steidel \etal\ (2003)
that lie in a $\approx$75~arcmin$^2$ region of the $\approx$2~Ms \cdfn\
(Alexander \etal\ 2003).  These sources have \hbox{$\cal R =$~23.3--25.6}
(median ${\cal R} = 24.9$), which are on average fainter than those in SSA22 by
$\approx$0.2~mag.  In total, 61 ($\approx$42\%) of the \cdfn\ LBGs have
spectroscopic redshifts from Steidel \etal\ (2003).  We constructed our field
LAE sample using the $z = 3.1$ LAEs from Gronwall \etal\ (2007) that were
within the extent of the \ecdfs\ \chandra\ coverage ($\approx$1008~arcmin$^2$
overlap), which consists of a central $\approx$2~Ms \chandra\ exposure (Luo
\etal\ 2008) that is flanked by four $\approx$250~ks (Lehmer \etal\ 2005a)
\chandra\ observations.  The Gronwall \etal\ (2007) sample of LAEs have
EW$_{\rm obs} > 80$~\AA\ and reach narrow-band depths of NB~$< 25.4$.
Spectroscopic follow up of 52 LAEs in the \ecdfs\ sample show that all of these
sources are at $z = 3.1$ (Gawiser \etal\ 2006; Gronwall \etal\ 2007).  In order
to make fair comparisons between field and protocluster LAEs, the Gronwall
\etal\ (2007) observational limits were used in our selection of the SSA22
sample discussed above.  Our field LAE sample consists of 257 LAEs.  In total,
our CDF field comparison samples consists of 403 unique sources at $z \sim 3$,
thus making the number of sources in the SSA22 and CDF field comparison samples
similar.

\subsection{X-ray Matching of LBG and LAE Samples}

We matched our SSA22 and CDF field comparison samples to the available
\chandra\ point-source catalogs (see $\S$~2 and references in $\S$~3.1).  For a
successful match, we required that the optical positions of our LBGs and LAEs
be offset by no more than 1\farcs0 from the \chandra\ source position.  Under
this criterion, we found \chandra\ counterparts for six LBGs and five LAEs in
SSA22.  One of these sources \hbox{J221739.04+001330.1} is both an LBG and LAE;
therefore, our SSA22 sample consists of ten unique \xray--detected sources at
$z \sim 3$ (see Table~1 for detailed properties).  We find that four of our six
SSA22 LBGs and all five SSA22 LAEs have spectroscopic redshifts.  In Figure~1,
we show the positions of our \xray--detected LBGs and LAEs.  We note that the
majority of our \xray--detected sources lie in high LAE density regions.  

We determined the expected number of false matches for our SSA22 sample by
shifting the 384 LBG plus LAE optical source positions by small offsets and
rematching them to our $\approx$400~ks \chandra\ point-source catalog.  We
performed eight such trials using positional shifts of 5\farcs0 and 10\farcs0
and found an average of $\approx$0.9 false matches per trial.  Based on this test, we
expect spurious matches to have a limited affect on our results.  

For our CDF field comparison sample, we found \xray\ detections for eight LBGs
in the \cdfn\ and two LAEs in the \ecdfs.  We find that six of the eight \cdfn\
LBGs and one of the two \ecdfs\ LAEs have spectroscopic redshifts.  One of the
\ecdfs\ LAEs \hbox{J033307.61$-$275127.0} is a $z=1.6$ interloper due to the
detection of the C~{\small III]} line at $\lambda$1909.  We note that such
interlopers are not common and that due to the strong C~{\small III]} emission
line from an AGN, it is not surprising that \hbox{J033307.61$-$275127.0} was
detected in the \xray\ band.  
Since the interloper fraction for
LAEs is very small both in the \ecdfs\ and SSA22 samples (see $\S$~3.1), we
have chosen to remove \hbox{J033307.61$-$275127.0} from our subsequent
analyses.  We have experimented with the inclusion of this source in our
analyses and find that this has no material affects on our overall results.

In Table~1, we summarize the basic \xray\ properties of the \xray\ detected
sources in our SSA22 and CDF field comparison samples.  For all sources, we
calculated rest-frame \hbox{2--8~keV} and \hbox{8--32~keV} luminosities using
observed-frame \hbox{0.5--2~keV} and \hbox{2--8~keV} fluxes, respectively, and
the best available value for the redshift.  For comparison purposes, we note
that the rest-frame \hbox{2--8~keV} luminosity $L_{\rm 2-8~keV}$ can be
converted to the more commonly utilized \hbox{2--10~keV} bandpass luminosity
$L_{\rm 2-10~keV}$ following $L_{\rm 2-10~keV} = \beta L_{\rm 2-8~keV}$, where
\hbox{$\beta \approx (10^{-\Gamma_{\rm eff} + 2} - 2^{-\Gamma_{\rm eff} +
2})/(8^{-\Gamma_{\rm eff} + 2} - 2^{-\Gamma_{\rm eff} + 2})$}.  For sources
without only limits on $\Gamma_{\rm eff}$, we find on average $\beta = 1.30 \pm
0.07$ (1$\sigma$ standard deviation).  Since our \xray--detected sources cover
the redshift range of $z \approx$~2.0--3.4, we made small multiplicative
corrections to observed frame fluxes to correspond to rest-frame
\hbox{2--8~keV} and \hbox{8--32~keV} luminosities.  These correction factors
were computed assuming a power-law spectral energy distribution (SED) with
photon index $\Gamma = 1.4$.  For our ten \xray--detected sources in SSA22, the
mean multiplicative correction factor and 1$\sigma$ standard deviation is $1.02
\pm 0.12$.

%
%
\begin{figure}
\centerline{
\includegraphics[width=9.5cm]{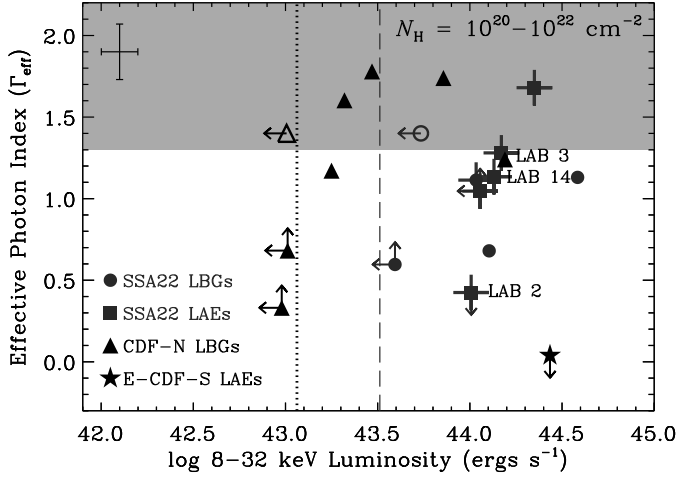}
}
\caption{
Effective photon index ($\Gamma_{\rm eff}$) versus rest-frame \hbox{8--32~keV}
luminosity for \xray--detected SSA22 and CDF field comparison samples.  The
shaded region indicates the range of $\Gamma_{\rm eff}$ values expected for a
source at $z = 3$ with intrinsic $\Gamma = 1.8$ and $N_{\rm H} =
10^{20}$--$10^{22}$~cm$^{-2}$.  SSA22 LBGs and LAEs have been indicated using
circles and squares, respectively; average 1$\sigma$ error bars for each
quantity have been indicated in the upper left-hand corner (see Table~1 for
individual values).  Sources that are spectroscopically confirmed to lie within
the protocluster have been highlighted with crosses; the location of LABs~2, 3,
and 14 have been noted (Geach et al. in preparation).  \cdfn\ field LBGs
and the \ecdfs\ LAE \hbox{J033316.86$-$280105.2} are shown as triangles and a
five-pointed star, respectively.  Open symbols indicate the two sources that
were detected only in the \hbox{0.5--8~keV} bandpass and therefore have an
adopted photon-index of $\Gamma_{\rm eff}=1.4$.  The vertical dashed and dotted
lines indicate the median sensitivity limit for {\it all} 382 and 403 $z\sim3$
sources in the SSA22 and CDF samples, respectively.  
}
\end{figure}

%
%
\begin{figure*}
\centerline{
\includegraphics[width=18.5cm]{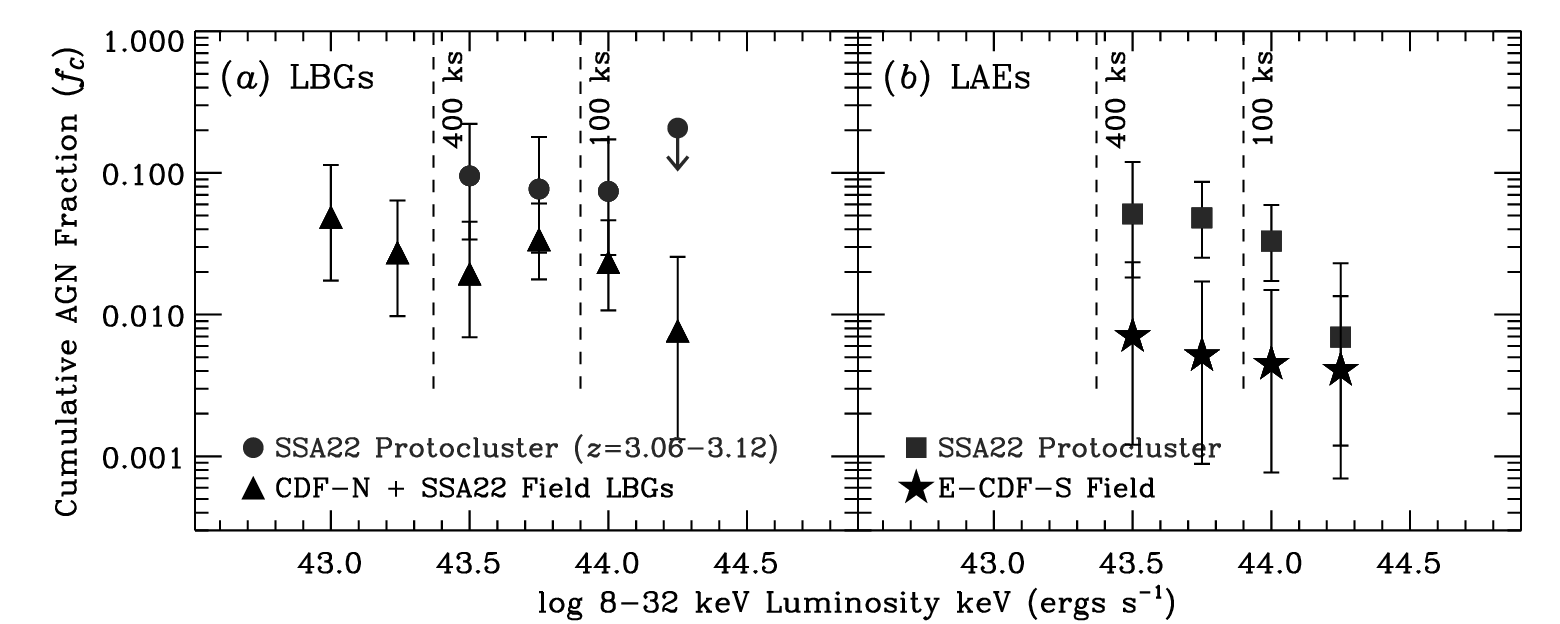}
}
\caption{
Cumulative fraction of galaxies harboring an AGN with rest-frame \hbox{8--32~keV}
luminosity larger than the given value (i.e., the AGN fraction, $f_C$) for
({\it a\/}) LBGs in the SSA22 protocluster ($z =$~3.06--3.12; {\it circles\/})
and \cdfn\ plus SSA22 field ({\it triangles\/}) and ({\it b\/}) LAEs in the
SSA22 protocluster ({\it squares\/}) and \ecdfs\ field ({\it five-pointed
stars\/}).  In both plots, the vertical dashed lines show the \chandra\ source
detection limits at $z = 3$ appropriate for 100~ks and 400~ks observations.
For both LBGs and LAEs, we find that $f_C$ is somewhat larger for the SSA22
protocluster compared to the field, suggesting either an increase in the
accretion activity (i.e., duty cycle and/or mean mass accretion rate) for
protocluster SMBHs or the presence of more massive SMBHs within protocluster
galaxies (see $\S$~4.3 and 5 for further discussion).
}
\end{figure*}

We expect that for highly-obscured sources, our rest-frame \hbox{2--8~keV}
luminosities may be significantly underestimated and would therefore be
considered as lower limits on the intrinsic luminosities.  However, we do not
expect such obscuration effects to have a significant influence on the
rest-frame \hbox{8--32~keV} luminosity, and we therefore use the rest-frame
\hbox{8--32~keV} luminosity to infer intrinsic energetics of \xray--detected
sources in our samples.  To identify signatures of absorption in
\xray--detected galaxies, we used basic observed \hbox{2--8~keV} to
\hbox{0.5--2~keV} band-ratios (BR\footnote{Here we define the band-ratio BR as
${\rm BR} = \Phi_{\rm 2-8~keV}/\Phi_{\rm 0.5-2~keV}$, where $\Phi_{\rm
0.5-2~keV}$ and $\Phi_{\rm 2-8~keV}$ are the net count-rates in the
\hbox{0.5--2~keV} and \hbox{2--8~keV} bandpasses, respectively.}) for
determining effective photon indices $\Gamma_{\rm eff}$.  The relationship
between band-ratio and $\Gamma_{\rm eff}$ was determined using {\ttfamily
xspec} (version 12.3.1; Arnaud~1996) to fit simple power laws, including only Galactic
absorption, to point sources in our \chandra\ catalog that were detected in all
three of our standard bandpasses (i.e., FB, SB, and HB) and had $>$50~counts in
the FB.  For these sources, we determined the emprical relationship between BR
and $\Gamma_{\rm eff}$ and used this relationship to compute $\Gamma_{\rm eff}$
for additional sources in our sample.

As discussed in $\S$~2, the most sensitive regions of the $\approx$400~ks
\chandra\ observations reach a $z = 3$ rest-frame \hbox{8--32~keV} luminosity
limit of $\approx$2.1~$\times 10^{43}$~\xlum.  This luminosity limit is
\hbox{$\simgt$20--50} times larger than the most luminous starburst galaxies in
the local universe (e.g., Persic \& Rephaeli~2007) and \hbox{$\simgt$10--40}
times more luminous than the expected \xray--emitting star-formation component
for $z\approx2$ ultraluminous submm-emitting galaxies (e.g., Alexander \etal\
2005).  We therefore conclude that all of our \hbox{$z\approx$~2.0--3.4}
sources detected in the observed \hbox{2--8~keV} bandpass are powered by AGNs.

\section{Results}

\subsection{X-ray Properties of LBG and LAE Samples}

In Figure~2, we show $\Gamma_{\rm eff}$ versus the logarithm of the
\hbox{8--32~keV} luminosity $\log L_{\rm 8-32~keV}$ for our SSA22 and CDF field
comparison samples.  We find that for sources detectable in the observed \hbox{2--8~keV}
bandpass of the SSA22 survey field ($\log L_{\rm 8-32~keV} \simgt 43.5$), we
have seven \xray--detected sources versus three in the CDFs.  These sources
cover \hbox{$L_{\rm 8-32~keV} \approx$~(3--50)~$\times 10^{43}$~\xlum} and have
values of \hbox{$\Gamma_{\rm eff} \simlt$~0.4--1.7} (median $\Gamma_{\rm eff}
\approx 1.1$), suggesting significant obscuring columns of \hbox{$N_{\rm H}
\simgt 10^{20}$--$10^{22}$~cm$^{-2}$}.  To test whether the protocluster
environment is influencing the observed absorption properties of the AGNs, we
performed \xray\ stacking (see Lehmer \etal\ 2008 for details) of five
\hbox{$z =$~3.06--3.12} SSA22 protocluster sources and five spectroscopically
confirmed field sources (in SSA22 and the CDFs) that had $\log L_{\rm 8-32~keV}
\simgt 43.5$ to determine mean photon indices.  We found stacked effective
photon indices of $\Gamma_{\rm eff} = 1.07^{+0.13}_{-0.12}$ and
$1.11^{+0.08}_{-0.07}$ for the protocluster and field, respectively.  This
suggests that on average, the \xray\ absorption properties of the most luminous
protocluster and field AGNs are similar.

\subsection{AGN Fraction and X-ray Stacking Results}

To assess whether the AGN activity per galaxy in the SSA22 protocluster
environment is quantitatively different from that observed in the field, we
compared the rest-frame \hbox{8--32~keV} luminosity dependent cumulative AGN
fractions, $f_C$, for LBGs and LAEs in the protocluster with those in the
field.  For a given sample of galaxies, $f_C$ can be computed following the
procedure outlined in $\S$~5.2 of Lehmer \etal\ (2008).  Briefly, we determined
$f_C$ by taking the number of candidate AGNs in a particular galaxy sample with
a rest-frame \hbox{8--32~keV} luminosity of $L_{\rm 8-32~keV}$ or greater
($N_{\rm AGN}$) and then dividing it by the number of galaxies in which we
could have detected an AGN with luminosity $L_{\rm 8-32~keV}$ ($N_{\rm gal}$).
$N_{\rm gal}$ can be computed by considering the redshift of each galaxy and
its corresponding sensitivity limit, as obtained from spatially varying
sensitivity maps appropriate for each \chandra\ observation (see Alexander
\etal\ 2003; Lehmer \etal\ 2005a; Luo \etal\ 2008; Lehmer \etal\ in preparation
for further details).

Using exclusively the 168 LBGs in our SSA22 and \cdfn\ samples that had
spectroscopic redshifts from Steidel \etal\ (2003), we computed $f_C$ for both
protocluster LBGs (27 sources) that were within the physical boundaries of the
SSA22 protocluster (\hbox{$z =$~3.06--3.12}) and field LBGs (141 sources) in
the SSA22 field (i.e., outside the protocluster redshift range) and the \cdfn.
As mentioned in $\S$~2, our \chandra\ observations cover only a small fraction
of the entire SSA22 protocluster extent and we therefore do not constrain
further the physical boundaries of the protocluster in the transverse
direction.  We experimented by comparing $f_C$ for protocluster and field LBGs
in the SSA22 region alone, and although not well constrained, the results are
consistent with those found by combining the SSA22 and \cdfn\ field samples.
For our LAEs, we computed $f_C$ for the SSA22 protocluster and \ecdfs\ field
samples.  

%
%

\begin{table}
\begin{center}
\caption{Data Used For Computing AGN Fraction $f_C$}
\begin{tabular}{cccccccc}
\hline\hline
 & \multicolumn{3}{c}{SSA22 Protocluster} & \multicolumn{3}{c}{CDF + SSA22 Field} &   \\
 $\log L_{\rm 8-32~keV}$ & \multicolumn{3}{c}{\rule{1in}{0.01in}} & \multicolumn{3}{c}{\rule{1in}{0.01in}} &  \\
 (\flux) & $N_{\rm AGN}$ & $N_{\rm gal}$ & $f_C$(\%) & $N_{\rm AGN}$ & $N_{\rm gal}$ & $f_C$(\%) & Enh$^a$  \\ 
\hline
\multicolumn{8}{c}{$z \approx$~2--3.4 Lyman Break Galaxies} \\
\hline
\vspace{0.05in}
43.50 &   2 &  21 &       9.5$^{+12.7}_{-6.1}$ &   2 & 103 &        1.9$^{+2.6}_{-1.3}$ &       4.9$^{+11.7}_{-3.9}$\\
\vspace{0.05in}
43.75 &   2 &  26 &       7.7$^{+10.2}_{-5.0}$ &   4 & 118 &        3.4$^{+2.7}_{-1.6}$ &        2.3$^{+5.8}_{-1.7}$\\
\vspace{0.05in}
44.00 &   2 &  27 &        7.4$^{+9.8}_{-4.8}$ &   3 & 128 &        2.3$^{+2.3}_{-1.3}$ &        3.2$^{+7.8}_{-2.4}$\\
\vspace{0.05in}
44.25 &   0 &  27 &                    $<$20.7 &   1 & 130 &        0.8$^{+1.8}_{-0.6}$ &                    $<$27.0\\
\hline
\multicolumn{8}{c}{$z = 3.1$ Ly$\alpha$ Emitters} \\
\hline
\vspace{0.05in}
43.50 &   2 &  39 &        5.1$^{+6.8}_{-3.3}$ &   1 & 142 &        0.7$^{+1.6}_{-0.6}$ &       7.3$^{+17.0}_{-6.2}$\\
\vspace{0.05in}
43.75 &   4 &  83 &        4.8$^{+3.8}_{-2.3}$ &   1 & 194 &        0.5$^{+1.2}_{-0.4}$ &       9.3$^{+16.9}_{-8.7}$\\
\vspace{0.05in}
44.00 &   4 & 121 &        3.3$^{+2.6}_{-1.6}$ &   1 & 223 &        0.4$^{+1.0}_{-0.4}$ &       7.4$^{+13.3}_{-6.9}$\\
\vspace{0.05in}
44.25 &   1 & 144 &        0.7$^{+1.6}_{-0.6}$ &   1 & 246 &        0.4$^{+0.9}_{-0.3}$ &        1.7$^{+5.7}_{-1.3}$\\
\hline
\end{tabular}
\end{center}
$^a$ Measured enhancement of the AGN fraction, where Enh~$\equiv f_C$(protocluster)/$f_C$(field) and Enh~$> 1$ indicates an elevation in the average AGN activity per galaxy in the SSA22 protocluster (see $\S$~4.2 for details).
\end{table}

In Table~2, we show the basic data used to compute $f_C$ for both the SSA22
protocluster and SSA22 plus CDF field comparison samples.  In Figure~3, we
present $f_C$ versus $L_{\rm 8-32~keV}$ for the above LBG (Fig.~3$a$) and LAE
(Fig.~3$b$) samples.  Error bars on $f_C$ are Poissonian and were computed as
double-sided 68.27\% (1$\sigma$) confidence intervals and 3$\sigma$ upper
limits on $N_{\rm AGN}$ following the methods described in Gehrels~(1986).  We
find that for both the LBGs and LAEs there is suggestive evidence for an
enhancement\footnote{Here enhancement is defined as a ratio of AGN fractions in
the protocluster versus outside the protocluster that is larger than unity
(i.e., $f_C$[protocluster]/$f_C$[field]~$> 1$).} of the AGN fraction for
\hbox{$\log L_{\rm 8-32~keV} =$~43.5--44.25} in the protocluster environment
versus the field.  For $L_{\rm 8-32~keV} \simgt 10^{43.5}$~\xlum, this
enhancement is measured to be $4.9^{+11.7}_{-3.9}$ and $7.3^{+17.0}_{-6.2}$ for
LBGs and LAEs, respectively.  Error bars on the enhancement are 1$\sigma$ and
were computed following the numerical error propagation method outlined in
$\S$~1.7.3 of Lyons~(1991).  To determine the significance of a measured
enhancement, we computed the integrated Poissonian probability for an overlap
between the protocluster and field AGN-fraction error distributions (i.e., the
probability that the protocluster and field AGN fractions are consistent).  For
the LBG and LAE AGN fraction enhancements, we found overlapping probabilities
of $\approx$24\% and $\approx$22\% (i.e., suggested enhancements at the
$\approx$76\% and $\approx$78\% significance levels), respectively.  As
discussed in $\S$~3.1, of the 27 LBGs in the SSA22 redshift spike, only 7
($\approx$26\%) are identified as also being LAEs.  We can therefore consider
these AGN fraction enhancement measurements for LBGs and LAEs to be two roughly
independent results, with a mean enhancement of $6.1^{+10.3}_{-3.6}$ and a
multiplicatively combined enhancement significance at the $\approx$95\%
confidence level.

To assess whether lower luminosity AGN and star-formation activity is enhanced
in the SSA22 protocluster over the field, we performed \xray\ stacking
following the procedure outlined in $\S$~4.2 of Lehmer \etal\ (2008).  We
restricted our stacking analyses to sources that were within 6\arcmin\
($\approx$2.7~Mpc at $z = 3.09$) of the \chandra\ aimpoints where the imaging
quality is best.  For both our LBG and LAE samples, we stacked separately
\xray--undetected galaxies that were within the SSA22 protocluster redshift
range and galaxies that were outside the protocluster including sources in the
SSA22 field itself and our CDF comparison fields.  For these four samples, we
detected significantly (i.e., at the $\simgt$3$\sigma$ level) the \xray\
emission from our field LBGs (non-spike SSA22 and \cdfn\ sources) and the SSA22
LAEs in the observed-frame \hbox{0.5--2~keV} band.  For the field LBGs, we find
a mean rest-frame \hbox{2--8~keV} luminosity of $(2.1 \pm 0.7) \times
10^{41}$~\xlum\ (4.4$\sigma$ significance), which is in good agreement with
that found from past stacking analyses of $z \approx 3$ LBGs (e.g., Brandt
\etal\ 2001; Nandra \etal\ 2002; Lehmer \etal\ 2005b, 2008; Laird \etal\ 2006).
For \xray--undetected LBGs in the SSA22 protocluster spike, we constrain the
mean rest-frame \hbox{2--8~keV} luminosity to be $\simlt$$1.3 \times
10^{42}$~\xlum\ (3$\sigma$ upper limit).  For our stacked LAEs, we constrain
the mean rest-frame \hbox{2--8~keV} luminosity to be $(6.9 \pm 2.9) \times
10^{41}$~\xlum\ (3.9$\sigma$ significance) and $\simlt$$4.9 \times
10^{41}$~\xlum\ for the SSA22 protocluster and \ecdfs\ field samples,
respectively.  These stacked luminosities suggest that on average LBGs and LAEs
are respectively $\simlt$6.1 and $\simgt$1.4 times more luminous in the
protocluster than in the field.  Such constraints suggest that low-level AGN
and star-formation activity allows for enhancement in the SSA22 protocluster at
a level consistent with that seen in more luminous AGNs (i.e.,
\hbox{$\approx$1.4--6.1}).  

If the \xray\ emission from the stacked LBGs are dominated by star-forming
processes and the \xray--SFR correlation at \hbox{$z \approx$~0--1.4} (from
Lehmer \etal\ 2008) is similar at $z \approx 3$, then we find \xray--derived
SFRs of $\simlt$341~\sfr\ and $\approx$57~\sfr\ for protocluster and field
LBGs, respectively (assuming a Kroupa~2001 initial mass function).  For
comparison, we utilized the LBG ${\cal R}$-band magnitudes provided by Steidel
\etal\ (2003) and an SED appropriate for LBGs (see $\S$~2.2 of Lehmer \etal\
2005b for details) to determine rest-frame UV luminosities and compute
UV-derived SFRs.  Following the \hbox{UV--SFR} relation from equation~1 of Bell
\etal\ (2003), we find average UV-derived SFRs of $\approx$11.2~\sfr\ and
$\approx$7.1~\sfr\ for protocluster and field LBGs, respectively.  If the
\xray\ emission is a reasonable tracer of the unobscured SFRs for these LBGs,
then the rest-frame UV emission from these sources would be obscured on average
by factors of $\simlt$30.4 and $\approx$8.0 for protocluster and field LBGs,
respectively.  We note that the obscuration factor for the field LBGs (i.e.,
$\approx$8.0) is $\approx$2 times larger than the average obscuration factor
measured using rest-frame UV spectral slopes for similar $z \approx 3$ LBGs
(e.g., Adelberger \& Steidel 2000), suggesting that \xray\ emission from lower
luminosity AGNs is likely contributing a significant fraction ($\approx$50\%)
of the stacked \xray\ signal for our LBG samples.

%
%
\begin{figure}
\centerline{
\includegraphics[width=9.5cm]{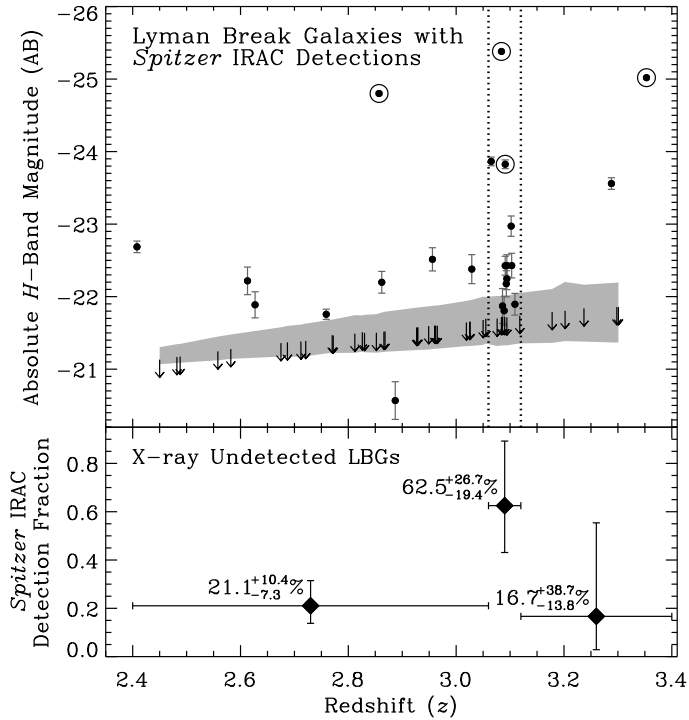}
}
\caption{
({\it top panel\/}) Rest-frame absolute $H$-band magnitude $M_H$ for LBGs in
our sample that were IRAC detected ({\it filled circles\/} with 1$\sigma$ error
bars); \xray--detected LBGs have been highlighted with open circles.  Upper
limits on $M_H$ are shown for IRAC-undetected LBGs, which were calculated using
SEDs that were fixed to 3.6$\mu$m photometery upper limits (5$\sigma$); $M_H$
upper limits calculated from SEDs with the median IRAC-detected best-fit
stellar age ($\approx$130~Myr) are shown as downward-pointing arrows and the
shaded region indicates the range of $M_H$ upper limits 
corresponding to the range of SEDs covering the quartile
range of IRAC-detected best-fit stellar ages ($\approx$~40--500~Myr).  For
reference, the redshift boundaries for the SSA22 protocluster at $z = 3.09$
have been outlined with vertical dotted lines.  
({\it bottom panel\/}) Fraction of \xray--undetected LBGs that were IRAC
detected in three redshift bins: \hbox{$z =$~2.4--3.06}, \hbox{$z
=$~3.06--3.12} (i.e., the SSA22 protocluster redshift range), and \hbox{$z
=$~3.12--3.4}.  Here the fraction of IRAC detected LBGs is largest in the
protocluster, indicating that the protocluster LBGs are likely more IR-luminous
(and correspondingly more massive) on average than field LBGs.
}
\vspace{-0.1in}
\end{figure}

\subsection{Rest-Frame Near Infrared Properties of LBGs}

To constrain the stellar content of SSA22 protocluster galaxies, we computed
rest-frame $H$-band luminosities for LBGs in our sample.  We presume that the
$H$-band luminosity provides a reasonable proxy for stellar mass; however, we
note that for a given $H$-band luminosity, galaxies with older stellar
populations will have intrinsically larger stellar masses.  We restricted this
analysis to the 87 LBGs in SSA22 that had spectroscopic redshifts from Steidel
\etal\ (2003) in the range of \hbox{$z =$~2.4--3.4}, a redshift range over
which photometric limits used for computing rest-frame $H$-band luminosities
are not expected to vary significantly (see below).  We constructed broadband
photometric SEDs for LBGs in our sample using
$U_n$, $G$, and ${\cal R}$ band photometry from Steidel \etal\ (2003), $J$ and
$K$ band photometry from the UKIRT Infrared Deep Sky Survey (UKIDSS; Lawrence
\etal\ 2007), $H$ band imaging from the UKIRT Wide-Field Camera (WFCAM)
obtained in service time, and \spitzer\ IRAC photometry in the 3.6, 4.5, 5.8,
and 8.0~$\mu$m bands (from the \spitzer\ archive; originally from GO project
30328).  To obtain reasonable estimates of the rest-frame $H$-band luminosities
for our LBGs, we used the SED fitting capability within the photometric
redshift code {\ttfamily HYPERZ}\footnote{See
http://webast.ast.obs-mip.fr/hyperz/ for details on {\ttfamily HYPERZ}.} to
obtain model SEDs (see details below).  When fitting our available photometric
data, we included only LBGs that were detected in at least one of the four IRAC
bands (hereafter, IRAC detected).  This ensured that our SEDs were well
constrained near the rest-frame $H$ bandpass, which corresponds to
observed-frame \hbox{$\approx$5--8~$\mu$m} for \hbox{$z \approx$~2.4--3.4}.
Furthermore, we visually inspected the UKIRT and IRAC images of all of our LBGs
to identify sources that were either confused by nearby sources or contained
low-significance artifacts.  In total, we were left with 23 galaxies
($\approx$26\% of our LBG sample) at \hbox{$z \approx$~2.4--3.4} with reliable
photometry that we used for computing rest-frame $H$-band luminosities.

Our SED models were derived from Bruzual \& Charlot (1993) assuming a single
star-formation epoch with an exponentially decaying star-formation history
(time constants  = 1, 2, 3, 5, 15 and 30~Gyr) and a Miller \& Scalo (1979) IMF;
the model grid spans ages of \hbox{0--20~Gyr}.  We utilized reddening curves
from Calzetti \etal\ (2000) and fit the extinction over the range of \hbox{$A_V
=$~0--2~mags}.  In the fitting process, we fixed the galaxy redshift to the
spectroscopic values given in Steidel \etal\ (2003).  For each of our 30
IRAC-detected LBGs, we convolved the best-fit template spectrum with the UKIRT
$H$-band filter function to approximate the rest-frame $H$-band luminosity.  We
checked to see whether the stellar populations of protocluster and field
galaxies were notably different by measuring the rest-frame colors (e.g., $U-V$
and $V-H$) of our best-fit SEDs.  We found no statistically significant
differences between the protocluster and field colors; however, due to limited
source statistics, we were unable to rule out mean color differences at levels
of $\Delta U-V \simlt 0.4$ and $\Delta V-H \simlt 0.5$ magnitudes.

In the top panel of Figure~4, we plot the absolute $H$-band magnitude $M_H$
versus redshift for LBGs with \hbox{$z =$~2.4--3.4}.  We note that the fraction
of LBGs that were IRAC detected is significantly larger for sources in the
protocluster than in the field.  In the bottom panel of Figure~4, we show the
IRAC detection fraction in three redshift bins at \hbox{$z =$~2.4--3.06}
($\approx$$21.1^{+10.4}_{-7.3}$\%), \hbox{$z =$~3.06--3.12}
($\approx$$62.5^{+26.7}_{-19.4}$\%), and \hbox{$z =$~3.12--3.4}
($\approx$$16.7^{+38.7}_{-13.8}$\%).  We find that the IRAC detection fraction
of LBGs is 3.0$^{+2.0}_{-1.3}$ times higher in the protocluster than in the
field implying there must be a significantly lower-luminosity (as measured in
the rest-frame $H$-band) and hence potentially lower-mass population of field
galaxies than protocluster galaxies.  

For sources that were not detected individually in any IRAC bands, we
calculated upper limits on $M_H$ by tying characteristic SEDs to the 5$\sigma$
photometric limit in the 3.6$\mu$m band, which provides the tightest
constraints on the rest-frame near-infrared emission.  To reasonably account
for variations in the median SEDs of these IRAC-undetected LBGs, we chose to
calculate a reasonable range of $M_H$ upper limits for these sources by using a
variety of SEDs spanning the quartile range of IRAC-detected best-fit stellar
ages (\hbox{$\approx$40--500~Myr}); these upper limit ranges are shown in the
upper panel of Figure~4.  

We find that for {\it all} LBGs (including those with only upper limits and
their upper limit ranges) without \xray\ detections, the median values of $M_H
= -21.9$ and $M_H \simgt$~$-21.9$ to $-21.3$ (accounting for the range of $M_H$
upper limits described above) for the protocluster (i.e., \hbox{$z
=$~3.06--3.12}) and field (i.e., \hbox{$z =$~2.4--3.06} and \hbox{$z
=$~3.12--3.4}), respectively.   These constraints imply that the typical
protocluster galaxy is \hbox{$\simgt$1.2--1.8} times more luminous in the
$H$-band for protocluster LBGs in comparison to field LBGs.  This also suggests that
protocluster LBGs are more massive than those found in the field.  

\section{Discussion}

The elevation in the SSA22 protocluster AGN fraction discussed in $\S$~4.2 is
plausibly expected due to (1) an increase in the accretion activity through
either more frequent accretion episodes and/or higher median SMBH accretion
rates in the high-density environments and/or (2) an increase in the typical
\xray\ luminosity of protocluster SMBHs resulting from the presence of more
massive host galaxies and SMBHs on average.  

Theoretical studies of the assembly and merger history of galaxies (e.g.,
Volonteri \etal\ 2003; Micic \etal\ 2007) suggest that SMBHs and their host
galaxies build up mass more quickly in high-density regions due to major-merger
activity.  Steidel \etal\ (1998) estimate that the SSA22 protocluster LBGs
reside in relatively massive dark matter halos of $\sim$$10^{12}$~$M_{\odot}$
per galaxy.  It is expected that galaxies in such massive halos will undergo a
peak in major-merger frequency at \hbox{$z \approx$~2--4}.  Therefore, in the
SSA22 protocluster these merger events may potentially be responsible for both
building up galaxy mass and funneling cold gas into the central SMBH and
fueling AGN activity at a higher frequency than in the lower-density field
environment.  

As previously discussed in $\S$~4.3, we have found initial evidence suggesting
that typical LBGs in SSA22 are \hbox{$\simgt$1.2--1.8} times more massive in
the protocluster compared to the field.  Similarly, detailed multiwavelength
studies of the $z = 2.3$ protocluster HS~1700$+$643, a cluster similar in
overall size and mass to SSA22, have found evidence suggesting the protocluster
galaxies have ages and stellar masses that are a factor of $\approx$2 times
larger than galaxies outside the protocluster (Shapley \etal\ 2005; Steidel
\etal\ 2005).  Given the relationship between host-galaxy and SMBH mass in the
local universe, it is reasonable to expect that a similar relationship will
hold at \hbox{$z \approx$~2--4} where massive SMBHs have been found to reside
in massive galaxies (e.g., Mclure \etal\ 2006; Peng \etal\ 2006; Alexander
\etal\ 2003).  This suggests that the elevated AGN fraction in the protocluster
compared to the field could be due to the presence of more massive and luminous
accreting SMBHs.  Qualitatively, we see similar behavior for star-forming in
general at \hbox{$z \approx$~0.2--2}, where the AGN fraction for a particular
\xray\ luminosity is observed to increase positively with galaxy stellar mass
(see, e.g., Fig.~5$b$ of Daddi \etal\ 2007 and Fig.~14$b$ of Lehmer \etal\
2008).  

We note that reasonable mass measurements for the SMBHs in our $z \sim 3$
galaxies is presently beyond the capabilities of modern instrumentation.
However, if we assume that (1) the enhanced AGN fraction in the SSA22
protocluster is primarily due to the presence of more massive SMBHs than in the
field and (2) the typical \xray\ luminosity $L_{\rm 8-32~keV}$ scales linearly
with SMBH mass (i.e., all the SMBHs are accreting at roughly the same fraction
of the Eddington limit), then we can crudely estimate the elevation in SMBH
mass by scaling up the luminosity dependence (i.e., by multiplying $L_{\rm
8-32~keV}$ by an elevation factor) of $f_C$(field) until we obtain reasonable
agreement between $f_C$ for the protocluster and field.  Based on Figure~3$a$,
where $f_C$(field) extends to lower values of $L_{\rm 8-32~keV}$ than
$f_C$(protocluster), we find that we would have to scale $L_{\rm 8-32~keV}$ for
$f_C$(field) by a factor of \hbox{$\simgt$3--10} to give consistent values of
the AGN fractions for the protocluster and field.  Under these assumptions,
this implies that the typical SMBH mass may be \hbox{$\simgt$3--10} times
larger in the protocluster compared to the field.  These constraints are
broadly consistent with those obtained for the elevation in galaxy stellar mass
(i.e., \hbox{$\simgt$1.2--1.8}), suggesting that both the SMBHs and their host galaxies may
be simultaneously undergoing significant fractional growth.  We note, however,
that these constraints are insufficient for distinguishing whether the mass
ratios between protocluster SMBHs and their host galaxies are consistent with
local relations.

Due to environmental effects, it is expected that the massive galaxies in the
SSA22 protocluster will quickly evolve onto the red-sequence and enter a passive
state with insignificant star-formation and black-hole growth.  Observational
studies of distant populations of massive galaxies ($\simgt$10$^{11}$~\msol)
have shown that stellar growth and AGN activity fall off dramatically between
\hbox{$z \approx$~2--3} and $z = 0$ (e.g., Papovich \etal\ 2006).  By $z = 0$,
it is expected that the SSA22 cluster would virialize and then resemble a local
rich cluster, where the SMBH growth in the cluster galaxies will have ceased to
levels lower that of the field population (e.g., Dressler \etal\ 1985; Martini
\etal\ 2007).  

\section{Summary and Future Observations}

Using a new ultra-deep $\approx$400~ks \chandra\ observation of the SSA22
protocluster at $z = 3.09$, we have investigated the role of environment on the
accretion of SMBHs.  Our key results are as follows:

\begin{itemize}

\item We have cross-correlated samples of 234 LBGs from Steidel \etal\ (1998)
and 158 LAEs from Hayashino \etal\ (2004) with our \chandra\ catalog and find a
total of ten \xray--detected sources at $z \approx 3$; at least six and
potentially eight of these are members of the protocluster.  These sources have
rest-frame \hbox{8--32~keV} luminosities in the range of \hbox{(3--50)~$\times
10^{43}$~\xlum} and a mean effective photon index of $\Gamma_{\rm eff} \approx
1.1$ suggesting significant absorption columns of \hbox{$N_{\rm H} \simgt
10^{22}$--$10^{24}$~cm$^{-2}$}.  

\item We have determined the rest-frame \hbox{8--32~keV} luminosity-dependent
AGN fraction for galaxies within the SSA22 protocluster at $z = 3.09$ and
compared it with that measured for $z\sim3$ galaxies in the field.  We found
that the fraction of $z = 3.09$ LBGs and LAEs harboring an AGN with $L_{\rm
8-32~keV} \simgt 10^{43.5}$~\xlum\ is a factor of \hbox{6.1$^{+10.3}_{-3.6}$}
times larger in the protocluster than in the field (see $\S$~4.2 for details).
We attribute the enhanced AGN fraction in the SSA22 protocluster to be
plausibly due to an increase in galaxy merger activity that would lead to an
increase in the SMBH accretion activity (e.g., more frequent accretion episodes
and/or higher median accretion rates) and/or an increase in the \xray\
luminosities of protocluster SMBHs due to the presence of more massive SMBHs
and host galaxies on average.

\item To differentiate between these two possibilities, we utilized
optical--to--mid infrared photometry to measure rest-frame $H$-band
luminosities of LBGs in our sample, which we expect to be a reasonable tracer
of stellar mass.  We found evidence suggesting that the stellar masses of LBGs
are \hbox{$\simgt$1.2--1.8} times more massive in the protocluster than in the field (see
$\S$~4.3 for details), and hence the larger AGN fraction most likely reflects
more massive SMBHs and associated host galaxies.

\end{itemize}

To constrain better the above results, future multiwavelength observations are
needed.  For example, in the SSA22 field, deeper and/or wider LBG and LAE
surveys would allow for the detection of further protocluster members; in the
Steidel \etal\ (2003) LBG catalog, $\approx$54\% of the sources do not have
spectroscopic redshifts, which includes two \xray--detected LBGs (see Table~1).
Yamada \etal\ (in preparation) have performed a wider LAE survey of SSA22
covering $\simgt$1~deg$^2$ scales, which has revealed additional high-density
regions within the $z = 3.09$ protocluster.  Expanding the multiwavelength data
set available to include such regions would enable more stringent constraints
to be placed on the enhancement of AGN activity as a function of global and
local environment at $z\sim3$.  This will place direct constraints on the
mechanism that causes SMBHs to grow.  Furthermore, analyses of additional
protoclusters similar to SSA22 that also contain deep \chandra\ and
multiwavelength data (e.g., HS~1700$+$643) could be combined with these data to
place more significant constraints on the influence of environment on AGN
activity in the high-redshift universe.

\acknowledgements

We thank the anonymous referee for their careful review and useful suggestions,
which have improved the quality of this manuscript.  We gratefully acknowledge
the financial support from the Science and Technology Facilities Council
(B.D.L., J.E.G.) and the Royal Society (D.M.A., I.R.S.).  Additional support
for this work was provided by NASA through Chandra Award Number SAO G07-8138C
(S.C.C., C.A.S., M.V.) issued by the Chandra X-ray Observatory Center, which is
operated by the Smithsonian Astrophysical Observatory for and on behalf of a
NASA contract.  This work is based in part on observations made with the
Spitzer Space Telescope, which is operated by the Jet Propulsion Laboratory,
California Institute of Technology under a contract with NASA.  Some of the
data reported here were obtained as part of the UKIRT Service Program.  UKIRT
is operated by the Joint Astronomy Centre on behalf of the Science and
Technology Facilities Council of the U.K.

%

%

\end{document}